\documentclass[twocolumn,showpacs,preprintnumbers,prl,floatfix]{revtex4-2}

\usepackage{graphicx}
\usepackage{epsfig}
\usepackage{bm}
\usepackage{amssymb,amsmath}
\usepackage{lipsum}
\usepackage{multirow}
\usepackage{nth} 
\usepackage{hyperref}
\hypersetup{colorlinks=true,linkcolor=blue,urlcolor=blue,citecolor=blue}

\def\4he{$^4$He}
\def\3he{$^3$He}
\def\cm3{cm$^{-3}$}

\begin{document}

\title{
Systematic-free limit on new light scalar bosons via isotope shift spectroscopy in Ca$^+$ 
}

\author{Timothy T. Chang$^1$, Bless Bah Awazi$^1$, Julian C. Berengut$^2$, Elina Fuchs$^{3,4}$, and S. Charles Doret$^1$\footnote{scd2@williams.edu}}
\affiliation{
$^1$Department of Physics, Williams College, Williamstown, MA  01267\\
$^2$School of Physics, University of New South Wales, Sydney, New South Wales 2052, Australia\\
$^3$Leibniz Universitat Hannover, Institute of Theoretical Physics, Appelstr. 2, 30167  Hannover, Germany\\
$^4$Physikalisch-Technische Bundesanstalt, Bundesallee 100, 38116  Braunschweig,Germany}
\date{February 20, 2024}

\begin{abstract}
\noindent We report a precise measurement of the isotope shifts in the $4^2$S$_{1/2} \rightarrow 3^2$D$_{3/2}$ electric quadrupole transition at 732~nm in $^{42,44,48}$Ca$^+$ relative to $^{40}$Ca$^+$ via high-resolution laser spectroscopy of co-trapped ions, finding measured shifts of 2,775,392,374.8(6.0), 5,347,679,835.1(5.9), and 10,003,129,115.1(5.7)\,Hz between $^{42,44,48}$Ca$^+$ and $^{40}$Ca$^+$, respectively.  When combined with prior measurements on the $4^2$S$_{1/2} \rightarrow 3^2$D$_{5/2}$ transition [\href{https://journals.aps.org/pra/abstract/10.1103/PhysRevA.100.022514}{Phys. Rev. A \textbf{100}, 022514 (2019)}] a King Plot analysis shows the data to be consistent with linearity below the level of parts per billion. 
This observed linearity, which is free of nuclear systematics,
improves
the previous isotope-shift based limits of Ca$^+$ for couplings of a beyond the Standard Model scalar boson to electrons and neutrons by a factor of 3.
Our new limit excludes part of the coupling range remaining for a new physics interpretation after accounting for one higher-order nuclear term in the nonlinear King plot of Yb/Yb$^+$. 
\end{abstract}

\maketitle
%
The Standard Model (SM) has long been a triumph of physics, explaining much of the universe as we know it and predicting a variety of phenomena such as neutral weak currents \cite{Gla61} and the Higgs boson \cite{EB64, Hig64, GHK64} which were later detected in accelerator-based measurements.\cite{ATLAS:2012yve,CMS:2012qbp}  And yet, several experimental observations and theoretical challenges strongly suggest that the SM is incomplete, for example failing to explain the matter-antimatter asymmetry in the universe \cite{Huet:1994jb,Gavela:1994dt,DK03} or to propose a satisfying candidate for Dark Matter~\cite{BH18}. New theories abound which might extend the Standard Model.  Laboratory measurements are making great strides towards exploring the vast parameter space of possibilities, with precision measurements in atomic and molecular (AMO) systems providing a powerful tool via measurements of parity violation as well as searches for permanent electric dipole moments (EDM), dark matter, dark energy, extra forces, and tests of CPT invariance and Lorentz symmetry~\cite{DDS17, SBD18, Antypas:2022asj}.

A guiding principle of most AMO-based probes of physics beyond the Standard Model (BSM) is that new physics may reveal itself as a perturbation to SM predictions as a consequence of the influence of a new particle or interaction.  While such perturbations typically lead only to tiny effects, clever experimental comparisons for which SM backgrounds can be made common-mode can highlight differential signals, as with EDM searches \cite{RCW03, AAD18} or clock comparisons which look for time-variation of fundamental constants \cite{Marion:2002iw,Safronova:2019lex, BBB21}.  Another example of this idea is isotope-shift (IS) spectroscopy, wherein a series of isotope shift measurements in multiple transitions amongst the same set of nuclei can be combined into a King plot~\cite{King63, King84}. Under the assumption that the IS is dominated by mass-shift and field-shift terms which may be represented as a product of nuclear-dependent and electronic-transition-dependent factors, such a plot is linear to first order in the SM. However, perturbations due to BSM effects could reveal themselves in the form of a breaking of King's linearity.  Recent work \cite{DOP17e,FFP17e,MTY17,BBD18e} has shown that tests of King's linearity in atomic systems offer the potential to achieve unprecedented sensitivity to intermediate-mass bosons $\lesssim 50$~MeV/$c^2$, such as the proposed 17~MeV/$c^2$ protophobic vector gauge boson which might resolve the anomaly measured in $^8$Be decays~\cite{FFG16e,FFG17e}.  With this as motivation, a flurry of recent effort \cite{MPE19e,Manovitz:2019czu,KPD19,KPD23,CHA20e,SMK20e, RRB21e,FBD22e,OSI22e,HAC22e,DYH24e2} has pushed the bounds of precision measurements of IS to place strong limits on BSM physics.

In the event that a nonlinearity is observed, atomic and nuclear theory is required to understand potential SM sources and calculate their size. In principle these effects may be subtracted from experimental observations to constrain BSM contributions. In practice, however, sufficient precision cannot yet be reached in such many-body calculations. The size of the sources of the SM nonlinearities, such as nuclear polarizabilities and quadratic mass- and field-shifts, have been estimated for several atomic systems \cite{FGV18,ADF21,viatkina23pra}. In \cite{BDG20e} it was shown that a `generalized King-plot' (GKP) analysis can mitigate the challenges associated with constraining new physics in the presence of SM nonlinearities. However in systems like Yb, with multiple sources of nonlinearity apparent at current experimental accuracy \cite{CHA20e,HAC22e,OSI22e}, the method cannot unambiguously set constraints on new physics.
As such, it is desirable to make measurements in a variety of atomic systems, and especially those with minimal SM sources of nonlinearity.   Ca is thus a prime candidate since its modest nuclear charge $Z$ gives it a much smaller quadratic-field-shift-contribution to the IS than Yb. \cite{FGV18}


In this Letter we report precision measurements of IS within the 4$^2$S$_{1/2} \rightarrow 3^2$D$_{3/2}$ 732~nm electric-quadrupole transition between $^{40}$Ca$^+$ and $^{42,44,48}$Ca$^+$.  Our measurements, in combination with our previous  729~nm 4$^2$S$_{1/2} \rightarrow 3^2$D$_{5/2}$ transition measurement \cite{KPD19, KPD23}, improve upon the precision obtained by Solaro \emph{et al.} \cite{SMK20e} by roughly a factor of three.  Further, from the $\sim$6~Hz precision achieved on both the 729 and 732~nm transition measurements we obtain a King plot which is consistent with linearity at better than the $10^{-9}$ level.   These measurements yield a new IS-derived constraint, not affected by SM systematics, on a possible interaction between electrons and neutrons mediated by a scalar boson beyond the SM.

\begin{figure}[h!]
\includegraphics[width=0.48\textwidth]{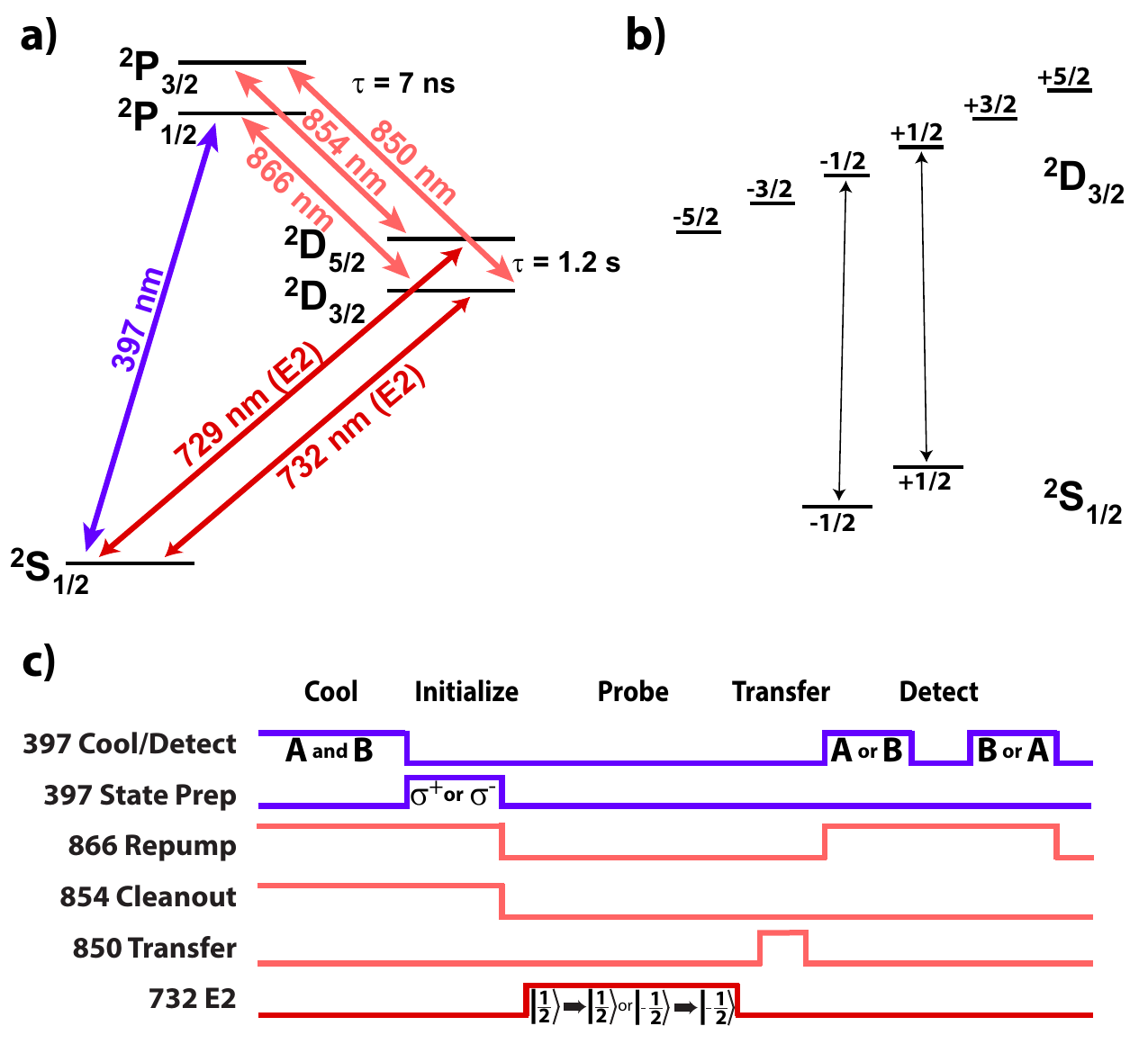}
\caption{\label{fig:levels_etc} (A) Electronic levels of Ca$^+$ with lifetimes and relevant transitions.  (B) Zeeman sublevels of the 732ñm transition.  (C) Time sequence of the experiment.  We alternate the order of probing $\Delta m=0$ transitions at 732~nm, and also the detection order (isotope A or B).}
\end{figure}

Our measurements proceed by direct spectroscopy of the 732~nm transition on a pair of co-trapped isotopes using an apparatus and measurement scheme similar to that described in \cite{KPD19}.  Briefly, isotope-selective photoionization \cite{LRH04e} via lasers at 423 and $\sim$375 nm is used to load ions of the desired species from a natural-abundance calcium oven source into separate minima of a double-well potential.  Additional lasers at 397 and 866~nm allow for Doppler cooling and fluorescence detection (Fig. \ref{fig:levels_etc}a).  A quantization field of approximately 1.75~G lifts the degeneracy of the ions' Zeeman states by a few MHz, enabling resolved spectroscopy of the magnetic sublevels (Fig. \ref{fig:levels_etc}b) on the 4$^2$S$_{1/2} \rightarrow 3^2$D$_{3/2}$ transition using a linewidth-narrowed laser locked to an ultrastable reference cavity.  


To measure the isotope shifts we first Doppler cool the ions, followed by state preparation of both co-trapped isotopes into one of the $|4^2$S$_{1/2}$,\,$m_j \!\!=\!\! \pm \frac{1}{2}\rangle$ states using $\sigma_\pm$ polarized light at 397~nm.  We then drive the 732~nm transition in both isotopes using frequency sidebands  generated with a fiber EOM referenced to a GPS-disciplined oscillator.  Simultaneously probing the two co-trapped isotopes dramatically reduces many possible sources of systematic errors, such as probe laser frequency drift or blackbody shifts, by making them common-mode.  Averaging interleaved detection of the $\Delta m_j = 0$ transitions from the two ground-state sublevels on msec timescales yields a measurement which is insensitive to differential first-order Zeeman shifts. 
Results of the measurement are determined using traditional electron shelving techniques.  However, since 
both the ground 4$^2$S$_{1/2}$ and excited $3^2$D$_{3/2}$ states are part of the ``bright" manifold of states associated with fluorescence measurement, an additional population-transfer step is required prior to detection.  We execute this transfer probabilistically by driving $3^2$D$_{3/2}$ population to the $4^2$P$_{3/2}$ state using lasers at 850~nm and rely upon spontaneous decay to $3^2$D$_{5/2}$ prior to fluorescence detection; see Fig. \ref{fig:levels_etc}c.  The transfer is thus incoherent, preventing the use of entanglement to enhance our measurement precision \cite{Manovitz:2019czu,RCK06e}, but the $\sim$6\% branching factor to $3^2$D$_{5/2}$ preserves adequate signal-to-noise to permit conventional spectroscopy techniques.   The measured 732-nm transition isotope shifts are shown in Table~\ref{table:Shifts}.  Quoted uncertainties (1$\sigma$) are statistics-dominated; systematic uncertainties of $\sim$2~Hz result principally from Stark shifts arising from uncompensateable micromotion~\cite{KPD19, BMB98e, BBD18e}.

\begin{table*}[t!]
	\begin{ruledtabular}
		\begin{tabular}{llll}
			& & \multicolumn{2}{c}{Isotope Shift (MHz)} \\
			Isotope pair $(A, A')$ & $\mu^{A,A'}$ ($10^{-6} \mathrm{u}^{-1}$)$^\textrm{a}$  & $^2$S$_{1/2}\rightarrow^2$D$_{3/2}$    & $^2$S$_{1/2}\rightarrow^2$D$_{5/2}$
 \\
            \hline
			(40, 42) &  1191.0358734(46) & 2 775.392 374 8(60)$^\textrm{b}$ &  2 771.872 467 6(76)$^\textrm{c}$  \\
			(40, 44) &  2274.305597(18) &  5 347.679 835 4(59)$^\textrm{b}$&   5 340.887 394 6(78)$^\textrm{c}$ \\
			(40, 46) &  3264.04178(17)& 7 778.302 524(2000)$^\textrm{d}$ &   7 768.401 000(2000)$^\textrm{d}$\\
			(40, 48) &  4171.5457614(93)& 10 003.129 115 1(57)$^\textrm{b}$ & 9 990.382 525 0(49)$^\textrm{c}$  \\
			
		\end{tabular}
	\end{ruledtabular}
\caption{\label{table:Shifts}Inverse-mass differences $\mu^{A,A'}$ and measured isotope shifts $\nu^{A,A'}$ for the five even Ca$^+$ isotopes. $\mu^{A,A'}$ is calculated from the atomic masses of the various isotopes less the electron masses and binding energies using data from \cite{WAK17e, NIST}.   Quoted uncertainties represent one standard deviation, and are dominated by statistical error. \\$^\textrm{a}$ Calculated from Ref. \cite{WAK17e, NIST} \hfill $^\textrm{b}$ This work. \hfill  $^\textrm{c}$ From Ref. \cite{KPD19, KPD23}.  \hfill $^\textrm{d}$ From Ref. \cite{SMK20e}.}
\end{table*}

Our measured isotope shifts can be interpreted with an eye towards a possible sum of SM and new physics contributions.  To start, consider the SM contributions to an isotope shift on some electronic transition denoted by its wavelength $\lambda$:
\begin{equation}
\label{eqn:IS}
\delta \nu^{A,A'}_\lambda \equiv \nu^{A}_\lambda - \nu^{A'}_\lambda = K_\lambda \mu^{A,A'} + F_\lambda \;\delta\langle r^2\rangle^{A,A'} + \ldots.
\end{equation}
Here $\mu^{A,A'} =\bigl(1/m_A - 1/m_{A'} \bigr)$ is calculated from the nuclear masses of isotopes with mass numbers $A,A'$ (Table \ref{table:Shifts}); $K_\lambda$ and $F_\lambda$ are the (linear) mass and field shift constants; and $\delta \langle r^2 \rangle^{A,A'}$ is the difference of the mean squared nuclear charge radii of the isotopes.  Ignoring higher-order terms, we can define a modified isotope shift (MIS) as 
\begin{equation}
    \label{eqn:MIS}
m\delta \nu^{A,A'}_\lambda \equiv {(\nu^A_\lambda - \nu^{A'}_\lambda)}/{\mu^{A,A'}}
\end{equation}
Canceling the $\delta \langle r^2 \rangle^{A,A'}$ term (which is the least well-measured) between two transitions $\lambda$ and $\lambda'$ yields the linear King's relation at leading order \cite{King63, King84}: 
\begin{equation}
m\delta \nu^{A,A'}_\lambda = K_\lambda - \frac{F_\lambda}{F_{\lambda'}} K_{\lambda'} + \frac{F_\lambda}{F_{\lambda'}} \;m\delta \nu^{A,A'}_{\lambda'}.
\end{equation}
A new boson $\phi$ beyond the SM leads to an additional term in eqn.~(\ref{eqn:IS}) \cite{BBD18e},
\begin{equation}
\label{eqn:KL}
\delta \nu^{A,A'}_\lambda = {K_\lambda}{\mu^{A,A'}} + F_\lambda \delta \langle r^2\rangle^{A,A'}  + \alpha_\textrm{NP} X_\lambda \gamma^{A,A'} + \ldots,
\end{equation}
where $X_\lambda$ characterizes the form of the $\phi$-particle's potential and its dependence on the transition $\lambda$,  $
\gamma_{A,A'} = A - A'$
indicates that $\phi$ couples to nuclei via the neutrons, and 
\begin{equation}
\alpha_\textrm{NP} = \frac{\bigl(-1\bigr)^s}{4\pi} \frac{y_e y_n}{\hbar c}
\end{equation}
describes the interaction of $\phi$ with electrons and neutrons in terms of its spin $s$ and coupling strengths $y_e, y_n$.
Such a new-physics contribution breaks the linear relationship.  As such, plotting the modified isotope shift of two transitions against one another and exploring the degree of linearity allows one to bound the size of $\alpha_{\text{NP}}$.

\begin{figure}[h]
\begin{center}
\includegraphics[width=0.48\textwidth]{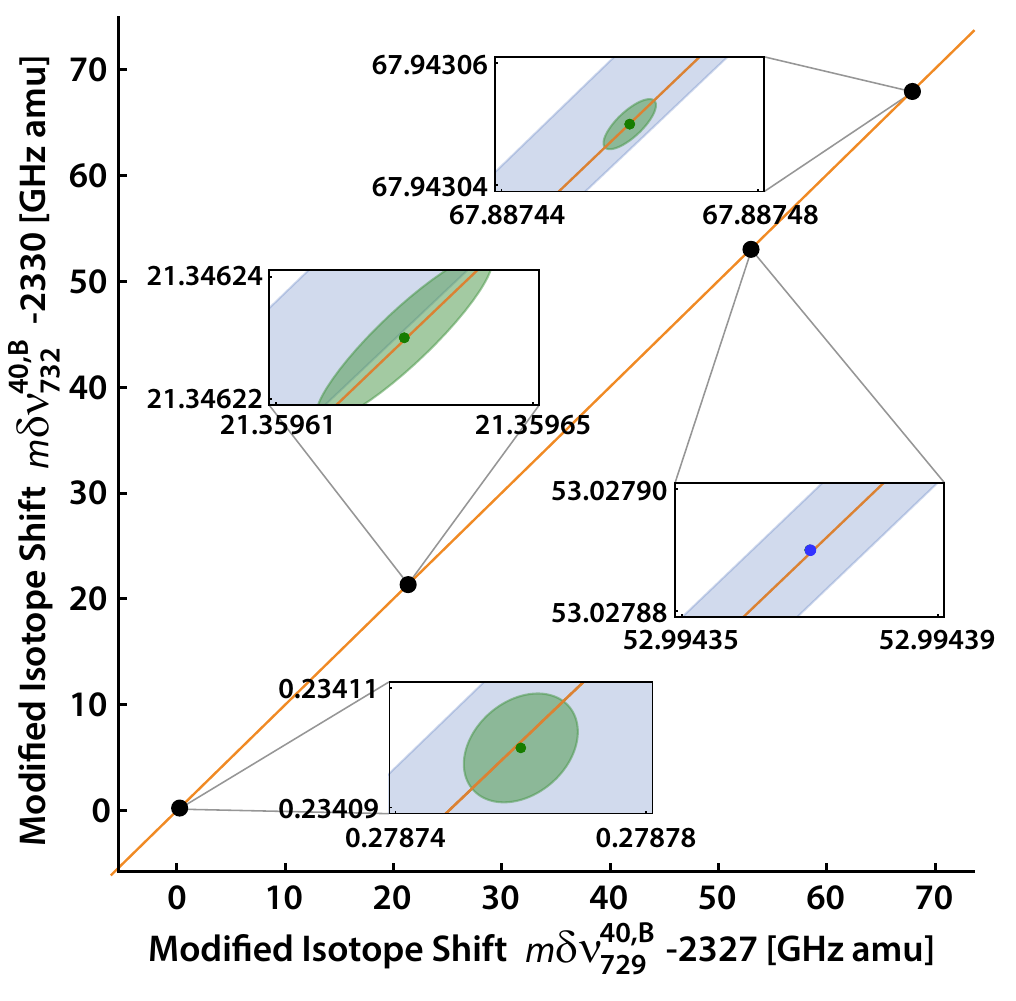}
\caption{\label{fig:KP} King Plot of the spin-zero isotopes of calcium at 732 and 729~nm, fit with a weighted 2D orthogonal distance regression.  Insets show uncertainty ellipses associated with this work in green, as well as prior measurements from \cite{SMK20e} in blue.  Measurement uncertainty parallel to the line of best fit is dominated by uncertainty in the measured nuclear masses, while that orthogonal to the line is limited by isotope shift measurements described here and in \cite{KPD19,KPD23}.}
\end{center}
\end{figure}

Figure \ref{fig:KP} represents a King plot of modified isotope shifts on the $4^2$S$_{1/2} \rightarrow 3^2$D$_{3/2, 5/2}$ transitions at $\lambda = 732$~nm, $\lambda' = 729~$nm  plotted against one another, including 1$\sigma$ error ellipses and a linear fit based upon a 2D weighted orthogonal distance regression~\cite{BBS87}. The highly-similar character of the two transitions considered here leads to the analysis not being limited by the precision of nuclear mass measurements \cite{SMK20e, CHA20e}; imprecision on the masses contributes a correlated uncertainty which lies almost entirely along the line of best fit.  This suppresses its impact by a factor of $2\times10^{-4}$, verified by Monte Carlo simulation.  Using eqn.~(\ref{eqn:KL}) above, we obtain the King plot parameters $K_{\lambda,\lambda '} = K_{732} - \frac{F_{732}}{F_{729}} K_{729} = -0.49630(3)$~GHz$\cdot$amu and $F_{\lambda,\lambda '} = \frac{F_{732}}{F_{729}} = 1.001483120(12)$.  We can also quantify the degree of nonlinearity in terms of the area spanned by the four points on the King plot, resulting in a significance of 0.42~$\sigma$ (see Appendix I).
The expected non-linearities in this King plot due to SM effects have been calculated and shown to be dominated by the second-order mass shift and nuclear polarization~\cite{viatkina23pra}. Unfortunately, the theoretically predicted nonlinearity is uncertain due to the strong cancellation of SM effects between the two transitions. Ref.~\cite{viatkina23pra} places an upper limit on the experimental accuracy needed to detect a nonlinearity at 200 Hz, which we are well below. Further refinement of the theory calculation is therefore needed in order to compare the nonlinearity expected from higher-order SM terms in Ca$^+$ to the present experimental precision.

\begin{figure}[t]
\begin{center}
\includegraphics[width=0.48\textwidth]{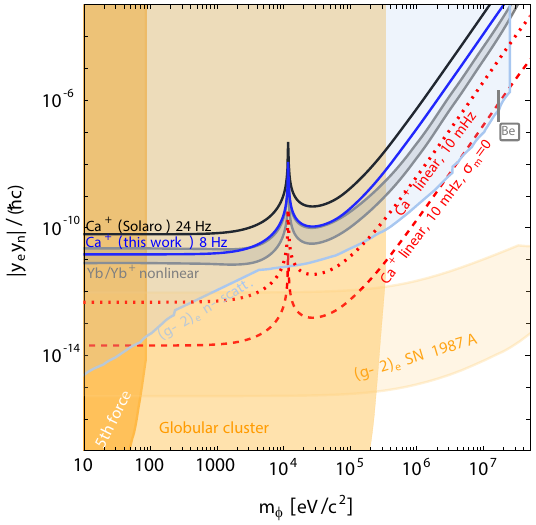}
\caption{\label{fig:DMexclusion} 
Constraints (2$\sigma$ upper bounds) on the strength of the coupling $|y_ey_n|/(\hbar c)$ between electrons and neutrons via a new boson $\phi$ with mass $m_\phi$.  The blue solid line represents the limit imposed by this work, where the 8~Hz precision is derived from Fig.~\ref{fig:KP}, roughly a quadrature sum of the two IS spectroscopic uncertainties.  The feature near $10^4$~eV/c$^2$ results from a transition-dependent resonance in the BSM sensitivity (see Appendix I for details)). 
  Plotted in black is the prior best result from Ca$^+$ by Solaro et al \cite{SMK20e} at $\sim$24\,Hz, while red shows projections based on a 10\,mHz IS measurement precision assuming no higher order systematic effects are found; the dashed curve includes only IS uncertainties, while the dotted curve incorporates current nuclear mass uncertainties. The gray band shows the bounds on IS measurements using Yb/Yb$^+$~\cite{HAC22e} assuming that the scalar boson is the dominant source of observed GKP nonlinearity. Additional shaded regions are constraints from other experiments such as the Casimir effect ~\cite{Bordag:2001qi,Bordag:2009zz}, the product $y_e y_n$ derived from from $(g-2)$ of the electron~\cite{Fan:2023gm2electronNorthwestern,Morel:2020dww} and neutron optics/scattering~\cite{Nesvizhevsky:2007by,Pokotilovski:2006up,Barbieri:1975xy,Leeb:1992qf,FFP17e}, star cooling in a Supernova \cite{Raffelt:2012sp} (note, however, its $\mathcal{O}(1)$ uncertainty~\cite{Blum:2016afe}), and as cooling in horizontal branch stars in the globular cluster~\cite{Grifols:1986fc,Grifols:1988fv}.  The protophobic boson model for the Beryllium anomaly is shown as a vertical bar. \cite{BBD18e,FFG16e,FFG17e}
}
\end{center}
\end{figure}

With a characterization of our agreement with King's linearity in hand, we can convert our measurement uncertainty into an upper-bound which constrains the strength of the coupling of a hypothetical boson $\phi$ to electrons and neutrons.  We take the coupling to arise from a Yukawa potential, 
\begin{equation}
V_\textrm{NP} = \alpha_\textrm{NP}\,\hbar c\,\frac{e^{-r m_\phi c/\hbar}}{r},
\end{equation}
where $m_\phi$ is the boson's mass.
The NP coefficients $X_\lambda$ defined in eqn.~(\ref{eqn:KL}) are as calculated in \cite{SMK20e} using AMBiT~\cite{kahl19cpc}; see \cite{SMK20e} for details.
We thus generate a new constraint on BSM bosons as indicated by the 2$\sigma$ upper bound plotted in solid blue in Fig.~\ref{fig:DMexclusion}, finding $y_e y_n / \hbar c < 1.4\times 10^{-11}$ in the low-mass limit.

Notably, our result excludes the coupling range needed to account for the nonlinearity detected in prior Yb$^+$~\cite{CHA20e} and Yb~\cite{FBD22e} measurements for the hypothesis of a new scalar boson being the dominant source.
This observation confirms that the nonlinearities observed in the Yb system are dominated by SM contributions. 
Our bound also excludes much of the new physics coupling range associated with the remaining nonlinearity of \cite{FBD22e} after accounting for one source of SM nonlinearity using the GKP. 
 
More recent measurements of Yb$^+$~\cite{HAC22e},  combined with the Yb measurements of~\cite{OSI22e}, use three transitions to take one source of SM nonlinearity into account via the GKP. Ref.~\cite{HAC22e} finds that the residual nonlinearity requires $6\times 10^{-12} \lesssim y_ey_n \lesssim 2\times 10^{-11}$ at the low-mass limit if a new scalar boson is assumed as the dominant source of this remaining nonlinearity (gray band in Fig~\ref{fig:DMexclusion}). Our limit from the linear Ca King plot excludes the upper part of that coupling range, but further studies are needed to fully differentiate between the hypotheses of a new boson or second SM term of higher order as the leading term causing the residual nonlinearity in~\cite{HAC22e}.

Given ongoing improvements in precision measurements of nuclear masses \cite{NGE20e} as well as the fact that the contribution of mass uncertainties to extracting limits from King plots is often strongly suppressed, there appears to be significant headroom for further improvement on these measurements. 
Measurement of the isotope shifts involving $^{46}$Ca at comparable precision to the more abundant isotopes would also be beneficial. 
Beyond that, established paths \cite{Manovitz:2019czu} exist to make IS measurements at the 10~mHz level; such improvements would allow IS-spectroscopy to break new ground in probing for fifth-force carriers in the range of $10^5 < m_\phi < 5\times 10^7$ eV/c$^2$. This offers a clear perspective to test -- as a target for King plot sensitivity -- the preferred coupling range associated with the proposed protophobic X17 boson \cite{FFG16e, FFG17e} at a mass of $m_X=17\,\rm{MeV}/c^2$, a potential explanation for the anomaly of nuclear Be$^*$ decays.   While it is expected that SM systematics due to higher-order mass- and/or field-shifts will complicate the interpretation of violations of King's linearity at this level of precision \cite{FGV18}, generalized approaches utilizing data from many transitions offer a path forward via techniques already applied in the Yb system~\cite{BDG20e, FBD22e}. Recent work with highly-charged calcium ions \cite{RRB21e} has paved the way towards precision IS-spectroscopy on forbidden M1 clock transitions in the charge states 11+ through 16+.  Besides adding more transitions to enable the generation of GKPs, these transitions feature orbitals which are dissimilar, substantially increasing their sensitivity to new physics~\cite{KSC18e}. Studies with additional atomic species would enable independent verification should evidence of a new boson emerge. Highly charged ions of Xe have recently been proposed~\cite{rehbehn23prl}, but barium may also be of particular interest.  Its greater mass offers substantially enhanced sensitivity to new physics, yet Ba still features modest, well-characterized sources of King plot nonlinearities from SM sources \cite{FGV18}, five naturally-occurring spin-zero isotopes, and two additional synthetic isotopes with $t_{1/2} > 1$~day which could be added to a King plot to further characterize any violations of linearity.   

In summary, we have reported measurements of the $^{40}$Ca$^+$-$^{42,44,48}$Ca$^+$ isotope shifts within the $4^2$S$_{1/2} \rightarrow 3^2$D$_{3/2}$ transition by direct spectroscopy of co-trapped isotopes at 732~nm with 6~Hz precision.  By combining these measurements with our group's past measurements on the parallel $4^2$S$_{1/2} \rightarrow 3^2$D$_{5/2}$ transition we generate a linear King plot and extract a new IS-based bound on the coupling strength of a fifth-force carrier $\phi$ to electrons and neutrons at the  $2 \sigma$ level in the low-mass limit of $y_e y_n / \hbar c < 1.4\times 10^{-11}$, free of limitations imposed by SM theory uncertainties which have appeared in similar work in other species.


\section{Appendix I: Limits on Coupling Strength}
\label{sec:Appendix}

To extract limits on the coupling strength of a new boson from our isotope shift measurements, we broadly follow the procedure presented in the Supplementary Material of Ref.~\cite{SMK20e}. We define vectors of modified isotope shifts~(\ref{eqn:MIS}) for each transition $\lambda$  as
\[
\boldsymbol{m\delta\nu}_{\lambda} = \left( 
    m\delta\nu_\lambda^{42,40}, m\delta\nu_\lambda^{44,40},
    m\delta\nu_\lambda^{46,40}, m\delta\nu_\lambda^{48,40} 
    \right)^T
\]
and $\boldsymbol{m\mu} = (1, 1, 1, 1)^T$. We then calculate the volume of a parallelepiped spanned by the three vectors $\boldsymbol{m\nu}_{\lambda}$, $\boldsymbol{m\nu}_{\lambda'}$, and $\boldsymbol{m\mu}$. King-plot linearity is equivalent to a vanishing volume, within experimental uncertainties.

Generally, to calculate the volume of parallelepiped formed from $N$-dimensional vectors $\boldsymbol{a}$, $\boldsymbol{b}$, and $\boldsymbol{c}$, we first define a $4\times2$ matrix $\boldsymbol{D} = \left( \boldsymbol{a}, \boldsymbol{b} \right)$. The projection of $\boldsymbol{c}$ onto the plane spanned by $\boldsymbol{a}$ and $\boldsymbol{b}$ is then
\[
\boldsymbol{p} = \left[ \boldsymbol{D}\cdot(\boldsymbol{D}^T\cdot\boldsymbol{D})^{-1}\cdot\boldsymbol{D}^T \right]\cdot\boldsymbol{c},
\]
and the parallelepiped volume is
\begin{equation}
\label{eq:vol}
V(\boldsymbol{a},\boldsymbol{b},\boldsymbol{c}) = \left| \boldsymbol{c} - \boldsymbol{p} \right|
\sqrt{\boldsymbol{a\vphantom{\boldsymbol{b}}}^2 \boldsymbol{b}^2 - (\boldsymbol{a}\cdot\boldsymbol{b})^2} .
\end{equation}
In the event of a nonlinearity caused by new physics, eqn.~(\ref{eqn:KL}) will cause the vector $\boldsymbol{m\nu}_{\lambda'}$ to not lie in the plane spanned by $\boldsymbol{m\nu}_{\lambda}$ and $\boldsymbol{m\mu}$ such that the parallelepiped volume is non-vanishing (the plane itself will also rotate). The deviation for each of the transitions is $\alpha_\textrm{NP} X_\lambda(m_\phi)\, \boldsymbol{m\gamma}$ with $\boldsymbol{\gamma} = (2, 4, 6, 8)$ in the current work and $\boldsymbol{m\gamma}$ defined by dividing each component by the corresponding $\mu^{A,A'}$. We can then extract the coupling strength according to
\begin{equation}
\label{eq:alphaNP}
\alpha_\textrm{NP} (m_\phi) = \frac{V(\boldsymbol{m\nu}_{\lambda},\boldsymbol{m\nu}_{\lambda'},\boldsymbol{m\mu})}{\left( X_{\lambda'}-\frac{F_{\lambda'}}{F_{\lambda}} X_{\lambda} \right) V(\boldsymbol{m\nu}_{\lambda},\boldsymbol{m\gamma},\boldsymbol{m\mu}) }
\end{equation}

To get the best possible limits on new physics, we combine the results of the measurements of this work with the 729~nm measurements presented in Ref.~\cite{KPD19} and $^{46,40}$Ca$^+$ measurements from \cite{SMK20e}. We obtain uncertainties using a Monte Carlo approach. We vary all experimental data assuming a Gaussian uncertainty distribution and for each sample we recalculate $\alpha_\textrm{NP}$ using (\ref{eq:alphaNP}). Repeating $10^5$ times gives us a stable uncertainty $\sigma_{\alpha_\textrm{NP}}$. The nonlinearity of the Ca$^+$ isotope shifts can be expressed as $\alpha_{NP}/\sigma_{\alpha_\textrm{NP}} = 0.42$. Note that the relatively large uncertainty in the absolute measurement of $^{46,40}$Ca$^+$ isotope shifts does not cause overly large errors because the difference in the fine-structure isotope shift -- which sets the uncertainty orthogonal to the line of best fit -- is measured to a precision of  21~Hz~\cite{SMK20e}.

The bounds plotted in Fig.~\ref{fig:DMexclusion} are the data value plus 2$\sigma$ uncertainty of $|y_e y_n| = 4\pi \alpha_\textrm{NP}$. The spikes in the sensitivity curves arise from cancellations of $X_{\lambda'}-\frac{F_{\lambda'}}{F_{\lambda}} X_{\lambda}$ in the denominator of (\ref{eq:alphaNP}). Note also that the curve marked ``Solaro'' is a re-calculation of the limits based on the data presented in~\cite{SMK20e} using our Monte Carlo uncertainty propagation.

\vspace{\baselineskip}

\begin{acknowledgments}

Thanks to Jacob Lezberg, Aidan Ryan, Ren\'ee DePencier Pi\~nero, and Sonya Dutton, each of whom contributed to the development of various aspects of the experimental apparatus.  This work was supported by the NSF under RUI grant numbers PHY-1707822 and PHY-2207957 and by a Cottrell Scholar Award from the Research Corporation for Science Advancement. EF acknowledges support by Germany’s Excellence Strategy – EXC-2123 QuantumFrontiers – 390837967.

\end{acknowledgments}

\bibliographystyle{apsrev}
\bibliography{reference_database}

\end{document}